\documentclass[twocolumn,aps,pra,showpacs,raggedbottom,nobalancelastpage,amssymb,superscriptaddress]{revtex4-1}

\usepackage{graphicx}
\usepackage{amsmath}
\usepackage{amssymb}
\usepackage{bm}
\usepackage[usenames]{color}
\usepackage{amsfonts}
\usepackage{appendix}
\usepackage{blindtext}

\usepackage{color,soul}

\newcommand{\ket}[1]{| #1 \rangle}
\newcommand{\bra}[1]{\langle #1 |}

\begin{document}
\newcommand{\comm}[1]{\hl{#1}}
\newcommand{\commb}[1]{{\sethlcolor{cyan}\hl{#1}}}

\title{Entanglement spectrum of mixed states}
\author{Evert van Nieuwenburg}
\affiliation{Institute for Theoretical Physics, ETH Zurich, 8093 Z{\"u}rich, Switzerland}
\affiliation{California Institute of Technology, Pasadena, CA 91125, USA}
\author{Oded Zilberberg}
\affiliation{Institute for Theoretical Physics, ETH Zurich, 8093 Z{\"u}rich, Switzerland}

\begin{abstract}
Entanglement plays an important role in our ability to understand, simulate, and harness 
quantum many-body phenomena.
In this work, we investigate the entanglement spectrum for open one-dimensional systems, 
and propose a natural quantifier for how much a 1D quantum state is entangled while being subject 
to decoherence. 
We demonstrate our method using a simple case of single-particle evolution and find 
that the open system entanglement spectrum is composed of generalized concurrence values, as well as 
quantifiers of the state's purity. 
Our proposed entanglement spectrum can be directly obtained using a correct scaling of 
a matrix product state decomposition of the system's density matrix.
Our method thus offers new observables that are easily acquired 
in the study of interacting 1D systems, and sheds light on the approximations employed in matrix product 
state simulations of open system dynamics.
\end{abstract}


\maketitle

\section{Introduction}
The wave nature of quantum-mechanical particles leads to fundamental physical implications such as interference and entanglement. 
The latter can be understood as a quantifier of how nonlocal a specific quantum state is, and has become a ubiquitous tool for understanding quantum many-body physics~\cite{Amico2008,Laflorencie2016,Ho2017}. Indeed, the study of entanglement plays an important role in research fields such as quantum information and quantum computation, which push technological advances towards the utilization of quantum mechanics in real-life applications~\cite{Horodecki2009}. 

Entanglement of a quantum mechanical state $\ket{\psi}$ can be considered with respect to a bipartition of the state into two parts $A$ and $B$.
Typically, the bipartitioning is taken as a spatial cut that divides the system into two equal halves. 
If the state can be decomposed into a product of state $\ket{\psi_A}$ in subsystem $A$ and a state $\ket{\psi_B}$ in subsystem $B$, the state $\ket{\psi}=\ket{\psi_A}\otimes\ket{\psi_B}$ is considered non-entangled.
Conversely, when the state can not be decomposed into a product with respect to the bipartitioning, it is entangled.
Note that these statements are made with respect to a given basis, i.e. a state can be non-entangled in one basis but entangled in another.
Additionally, a given state can be entangled for one bipartition but not for another. 
 
One method of quantifying the entanglement of a pure state with respect to a given bipartitioning, is using the entanglement entropy $S$. 
To compute it, one considers the density matrix $\rho = \ket{\psi}\bra{\psi}$ corresponding to the state $\ket{\psi}$. 
The bipartitioning into parts $A$ and $B$ is obtained through a partial trace s.t. $\rho_A = \textrm{Tr}_B \rho$. 
The reduced density matrix $\rho_A$ contains the full information of the state within subsystem $A$. 
Additionally, it stores information on the amount of entanglement that exists between subsystems $A$ and $B$. 
Namely, one can interpret the reduced density matrix as $\rho_A = e^{-H_e}$, where $H_e$ is known as the entanglement Hamiltonian~\cite{Li2008}. 
The eigenvalues $\epsilon_i$ of $H_e$ are known as the entanglement spectrum (ES). 
Typically, one works with the eigenvalues $\lambda_i=e^{-\epsilon_i}$ of $\rho_A$ directly, and then computes the entanglement entropy $S\equiv -\textrm{Tr}\rho_A \log \rho_A =-\sum_i \lambda_i \log \lambda_i$ as a measure of entanglement. If $S=0$ the state is non-entangled; non-zero $S$ signals entanglement.

\begin{figure}[ht!]
\begin{center}
\includegraphics[width=0.8\columnwidth]{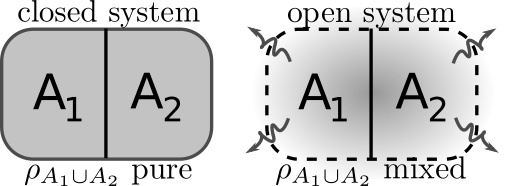}
\end{center}
\caption{Entanglement between subsystems $A_1$ and $A_2$. On the left, the full system $A = A_1 \cup A_2$ is decoupled from an environment and hence in a pure state. On the right the system was coupled to a traced-out environment, and therefore generally in a mixed state. In this work we focus on entanglement between the two subsystems, $A_1$ and $A_2$, when the system as a whole can decohere due to the presence of an environment.}
\label{fig:systemandenvironment}
\end{figure}

The reduced density matrix generally describes a mixed state~\cite{Nielsen2011}, i.e. its purity $\mathcal{P}_A\equiv {\rm Tr}\left\{\rho_A^2\right\}$ is less than unity. 
This procedure of tracing out a subsystem is a natural description of open quantum systems, in which the traced out part is the environment. In this work we consider entanglement between subsystems $A_1$ and $A_2$ of an open quantum system (cf.~Figure~\ref{fig:systemandenvironment}), where the latter may initially be non-pure due to the tracing out of an environment $B$. Due to the linearity of the trace operation, this is in principle equivalent to considering $\rho_{A_1} = \textrm{Tr}_{A_2 \cup B} ( \rho_{A_1 \cup (A_2 \cup B)} )$, representing a different bipartitioning of the whole system. Nevertheless, considering the route of first tracing out $B$ provides a physical interpretation of the resulting entanglement properties.

Here we investigate a natural definition for the entanglement spectrum of open systems that can serve as a direct measure for entanglement and correlations between two subsystems $A_1$ and $A_2$. 
We perform a bipartitioning of the mixed state of an open system by interpreting the corresponding density matrix as a vector (i.e. by vectorizing it), and analyze the obtained generalization of the entanglement spectrum for mixed states~\cite{Zwolak2004, Verstraete2004}, referred to as the operator space entanglement spectrum (OSES)~\cite{Prosen2007, Zhou2017}. 

We find here that the OSES directly encodes information about the purity of the mixed state as a whole, as well as of its bipartitioned subparts. 
Focusing on the case of a single-particle excitation we highlight that the OSES additionally encodes nonlocal bipartite entanglement information (in the form of generalized concurrence values), similar to the closed system entanglement spectrum. 
Importantly, the OSES can be readily obtained numerically using matrix product state (MPS) decomposition of the system's vectorized density matrix. 
As result, our work highlights what kind of information is neglected when using such algorithms with a finite bond dimension, and appropriate transformations before truncation may be used to preserve these quantities~\cite{White2017}.

Our procedure for defining the OSES of the mixed state operator $\rho_A$ is identical to that used in the discussion of operator space entanglement entropy (OSEE)~\cite{Prosen2007, Pizorn2009, Zwolak2004}. The procedure for numerically obtaining the OSES differs in an important way from the OSEE procedure by a re-scaling factor. The OSES has rarely been considered directly, but was previously shown to indicate symmetry protected topological states in open systems via its degeneracy structure~\cite{Nieuwenburg2014}, by a direct extension of the known results for the closed system~\cite{Pollmann2010}.

Our paper is structured as follows: in Sec.~\ref{sec:OSES}, we review the formalism for obtaining the OSES. In Sec.~\ref{sec:singleexcitationOSES}, we obtain an analytic description for the case of single-excitation mixed states, followed by two demonstrations of these results for simple one-dimensional spin-chains on small chains in Sec.~\ref{sec:demos}. We discuss the impact and scope of our results in Sec.~\ref{sec:outlook}.

\section{Operator space entanglement spectrum}
\label{sec:OSES}
Inspired by the process of obtaining the entanglement spectrum at a bipartitioning of a pure state~\cite{Li2008}, we present here a method for obtaining a meaningful entanglement spectrum for a mixed state. 
We refer to Appendix~\ref{app:singleparticleES} for more details on the pure state entanglement spectrum. 

We consider the density matrix $\rho$ of an open system $A$, e.g. $\rho$ may be obtained by tracing out an environment. The density matrix can be written in a Liouville space as a vector $\ket{\rho}$~\cite{Zwolak2004, Verstraete2004} by a procedure referred to as vectorization.
Then, in analogy to the construction of density matrices from pure states, we can define a Hermitian matrix $Q$ as 
\begin{align}
	Q=\ket{\rho}\bra{\rho}\,.
\end{align}	

We now consider a bipartition of $Q$ into subparts $A_1$ and $A_2$, and trace out the latter to obtain
\begin{align}
	Q_{A_1} = \textrm{Tr}_{A_2} Q  \, .
\end{align}	
We can write $Q_{A_1} = e^{-\tilde{H}_e}$, where $\tilde{H}_e$ is interpreted as some Hamiltonian with a spectrum $\tilde\epsilon_i$, which is the operator space entanglement spectrum. 
In the following, we will work with the eigenvalues $\Lambda_i=e^{-\tilde\epsilon_i}$ of $Q_{A_1}$ and refer to these exponential values as the OSES. 

The main aim of our work is to investigate the physical information encoded in the $\Lambda_i$.
Before presenting a more rigorous analysis of the OSES, we want to highlight one immediate consequence of this definition of $\Lambda_i$.
From the structure of $Q$ it is apparent that the trace over the matrix $Q$ is $\text{Tr} Q = \text{Tr} \rho^2$, i.e., it equals the purity of the system. 
Considering that (i) $\Lambda_i$ are the eigenvalues of $Q_{A_1}$ and (ii) $\text{Tr} Q= \textrm{Tr}_{A_1} Q_{A_1}$, we readily obtain that the sum of the OSES is the purity $\text{Tr} \rho^2=\sum_i \Lambda_i$. 
This important property is sometimes overlooked in matrix product state simulations of open systems, because the $\Lambda_i$ are typically scaled s.t. $\sum_i \Lambda_i^2 = 1$.

\subsection{Partial trace of $Q$}
Let us now continue with the partial-tracing of $Q$. 
We begin by introducing a set of local basis matrices that span a local Hilbert space $\mathcal{H}$, and denote them $\Gamma^{[n]}_{i_n}$. The superscript $n$ indicates the site $n$, and the subscript $i_n = 1\ldots d^2$ with $d = \textrm{dim} \mathcal{H}$. A general density matrix $\rho$ on $L$ sites can now be decomposed using a tensor product of these basis matrices as:
\begin{align}
\rho = \sum_{i_1 \ldots i_L} P_{i_1 \ldots i_L} \Gamma^{[1]}_{i_1} \otimes \cdots \otimes \Gamma^{[L]}_{i_L} \,.
\label{eq:fanoForm}
\end{align}
In doing so, we have assumed that every site has the same local Hilbert space of dimension $d$. 
We then denote the vectorized matrices $|\Gamma^{[n]}_{i_n}\rangle$, and they are chosen to obey the following on-site normalization condition:
\begin{eqnarray}
	\langle \Gamma^{[n]}_{i_n} \mid \Gamma^{[n]}_{i_m} \rangle = \frac{1}{d}\textrm{Tr} \; \Gamma^{[n]\dagger}_{i_n}\Gamma^{[n]}_{i_m} = \delta_{i_n,i_m},
\label{eq:gammaorthonormal}
\end{eqnarray}
where the first equality defines the inner product between vectorized matrices. In the following, we will assume that the local Hilbert space is that of a spin-$1/2$, i.e. the local basis can be composed of the identity matrix plus Pauli matrices, $\sigma_{i_n}$ with $i_n=0,1,2,3$. In general for spins $S$, the generators of SU(2S+1) provide such a basis. We will introduce our choice for the basis matrices in the next section. For ease of notation, we will drop the superscript site labels wherever possible.

Using the vectorized density matrix in this basis, the matrix $Q$ can be written as follows:
\begin{align}
 \resizebox{.9\hsize}{!}{$\displaystyle Q  = \sum_{i_1\ldots i_L,\alpha_1\ldots\alpha_L} P_{i_1\ldots i_N} P_{\alpha_1\ldots\alpha_N} \ket{\Gamma_{i_1}}\cdots \ket{\Gamma_{i_L}}\bra{\Gamma_{\alpha_1}}\cdots\bra{\Gamma_{\alpha_L}} \,,$}
\end{align}
The process of tracing out the sites $m + 1$ through $L$ then leads to
\begin{align}
\label{eq:partialtracedQ}
 &\resizebox{.9\hsize}{!}{$\displaystyle\text{Tr}_{m+1\ldots L} Q = \sum_{j_{m+1}\ldots j_{L}} \bra{\Gamma_{j_{m+1}}} \cdots \bra{\Gamma_{j_L}} Q \ket{\Gamma_{j_{m+1}}}\cdots\ket{\Gamma_{j_L}}$} \\
  &\resizebox{.9\hsize}{!}{$\displaystyle = \sum_{i_1\ldots i_m,\alpha_1\ldots\alpha_m} C_{i_1\ldots i_{m};\alpha_1\ldots\alpha_{m}}\ket{\Gamma_{i_1}}\cdots \ket{\Gamma_{i_m}} \bra{\Gamma_{\alpha_1}}\cdots\bra{\Gamma_{\alpha_{m}}}$} \,,\nonumber
\end{align}
where we have defined the prefactors
\begin{multline}
\label{eq:C}
C_{i_1\ldots i_{m};\alpha_1\ldots\alpha_{m}}
=\\\left( \sum_{j_{m+1} \ldots j_{L}} P_{i_1\ldots i_{m};j_{m+1}\ldots j_L} P^*_{\alpha_1\ldots\alpha_{m};j_{m+1}\ldots j_L} \right)\,,
\end{multline}
which can be placed in a matrix $C$ of size $d^{2m}\times d^{2m}$. 
When considering a local basis composed of Pauli matrices for a spin-$1/2$ system, $C$ is therefore of size $4^m\times 4^m$.

Having defined a procedure for obtaining the partial-trace of $Q$, the OSES $\Lambda_i$ is obtained by diagonalizing the remaining matrix $Q_{A_1}$. 
This is equivalent to obtaining the eigenvalues of the matrix $C$. This is a challenging task in general, and we will focus on the case of a single excitation in the following. We note that efficient numerical methods for obtaining the OSES in the case of many-body systems exist, see Appendix.~\ref{app:MPS}.

\section{Single excitation}
\label{sec:singleexcitationOSES}
In this section, we detail a specific example for the type of information encoded in $\Lambda_i$.
We restrict ourselves to the case of a single excitation, and do not consider particle-loss processes here. Rather, we will include general decoherence processes that make the state non-pure.

As a set of basis matrices, we choose the set 
\[ \Gamma^{[n]}_{i_n} \in \left\{\begin{pmatrix} 1 & 0 \\ 0 & 0 \end{pmatrix}, \begin{pmatrix} 0 & 0 \\ 0 & 1 \end{pmatrix}, \begin{pmatrix} 0 & 0 \\ 1 & 0 \end{pmatrix}, \begin{pmatrix} 0 & 1 \\ 0 & 0 \end{pmatrix}\right\}, \]
and introduce the shorthand notation $\{\sigma_e, \sigma_f, \sigma_+, \sigma_-\}$ for these matrices, respectively. The matrices $\sigma_{e,f}$ describe an empty or full site, and $\sigma_{\pm}$ describe the corresponding off-diagonal elements. Notice that with this definition of the basis matrices, the $1/d$ normalization factor in Eq.~\ref{eq:gammaorthonormal} is not needed.

Using this notation, the basis states for the $L$-site system with a single excitation can be divided into two groups. 
The first group are basis states that are made out of products of $\sigma_{e,f}$, corresponding to the diagonals of the density matrix. In particular, for a single excitation this means only those states in which a single site $n$ has $\sigma_f$ as its local basis, whilst the others are empty with $\sigma_e$. The other group of basis states spans the off-diagonal elements (the coherences) of the density matrix and consists of states where a pair of sites are connected using $\sigma_{\pm}$ and the others are empty. These are the only allowed basis states for a single excitation, and we use the shortcut notation for the coefficients in Eq.~\eqref{eq:fanoForm} of the former type as $P_{j=f}$ (corresponding to e.g. $P_{e\ldots efe\ldots e}$ with the $f$ at position $j$), and the latter as $P_{j=+,l=-}$.

Since we will be considering mixed states, it is useful to decompose the $i,j$-entry of the density matrix as $\xi_{ij} \rho_{ij}$, where $\xi_{ij}$ are complex numbers representing decoherence, i.e., $\xi_{ii} \equiv 1$ on the diagonal of the density matrix and $\left|\xi_{ij}\right| \leq 1$ for the off-diagonal elements. In particular, a pure state has $\xi_{ij} = 1$ for all $i,j$.

In the above notation the coefficients $P_{j=f}$ correspond simply to $\rho_{jj}$, i.e. to the density $n_j$ at site $j$. Additionally, it sets an important relation for the case of an initially pure state, namely,
\begin{align}
	P_{j=+,l=-} P_{j=-,l=+} = \left|\xi_{jl}\right|^2 P_{j=f}P_{l=f}\,.
	\label{coh_to_dens}
\end{align}
This relation stems from the fact that the density matrix is a Hermitian operator that describes a pure state that can decohere. Since the density matrix is Hermitian, the left-hand side of Eq.~\eqref{coh_to_dens} can be further reduced to $|P_{j=+,l=-}|^2$.

We are now in a position to use these relations to obtain an exact expressions for the OSES. 
To do so, we construct the matrix $C$ obtained by a bipartitioning of the system between sites $m$ and $m+1$ [cf.~Eq.~\eqref{eq:C}] and obtain its eigenvalues $\Lambda_i$.
In the single excitation case, the matrix $C$ decomposes into three subblocks corresponding to the cases where: (i) the particle is fully traced out (i.e. it was in subsystem $A_2$ and no coherences connected it to subsystem $A_1$), (ii) the particle is fully on the remaining side (subsystem $A_1$), and (iii) the particle has some coherent part that spans across the bipartition. The OSES will be composed of the eigenvalues of these subblocks, which we discuss in detail in the next sections.

\subsection{The particle is fully in subsystem $A_2$}
\label{block1}
In this case the subblock consists of just a single element. 
The indices of $C$ correspond to empty chains due to the fact that the particle remained in the subsystem that was traced out, i.e., the indices of $C$ are $i_k = e$ and $\alpha_k = e$ for all $1\leq k \leq m$. Therefore, the OSES that originates from this block is directly the $C$-matrix element itself, i.e.,
\begin{eqnarray}
\label{eq:val1}
\Lambda_1 &\equiv& C_{\mathbf{e}_{m}\,;\,\mathbf{e}_{m}} \nonumber \\
           &=& \sum_{j=m+1}^{L} | P_{j=f} |^2 + 2 \sum_{j=m+1}^{L}\sum_{l=m+1,l\neq j}^{L} | P_{j=+,l=-} |^2\nonumber\\
	   &=& \sum_{j,l=m+1}^{L} |\xi_{jk}|^2 P_{j=f}P_{l=f}\equiv \mathcal{P}_{A_2} \,,
\end{eqnarray}
where we have used notation $\mathbf{e}_{m}\equiv \underbrace{e\ldots e}_{\textrm{m times}}$. We have, additionally, inserted the shorthand single-particle basis notation into Eq.~\eqref{eq:C}, and have used Eq.~\eqref{coh_to_dens}. Importantly, the last equality highlights that the expression \eqref{eq:val1} that we have obtained encodes the purity of subsystem $A_2$, i.e., $\mathcal{P}_{A_2}\equiv {\rm Tr} \rho_{A_2}^2$ with $\rho_{A_2}\equiv {\rm Tr}_{A_1} \rho$. One of the single excitation OSES' entries therefore corresponds to the purity of the traced-out subsystem.

\subsection{The particle is fully in subsystem $A_1$}
\label{block2}
In this case, the non-zero elements of $C$ have the form
\begin{align}
	C_{i_1\ldots i_m ; \alpha_1 \ldots \alpha_m} = P_{i_1 \ldots i_m; \mathbf{e}_{L-m}} P^*_{\alpha_1 \ldots \alpha_m; \mathbf{e}_{L-m}}\,,
\end{align}
representing the case where the single particle and its basis decomposition in the $A_1$ subsystem solely appear in independent $C$ elements. The size of this block is equal to the number of possible single particle configurations within the basis $i_1 \ldots i_m$, i.e., all the density and coherences within subsystem $A$. There are $m$ of the former, and $m(m - 1)$ of the latter, and hence the block is of size $m^2 \times m^2$.

The subblock of this case can be directly expressed as $\mathbf{v}\cdot \mathbf{v}^\dagger$, where $\mathbf{v}$ is a column vector with elements $P_{i_1 \ldots i_m; \mathbf{e}_{L-m}}$ as its entries. Correspondingly, there is a single eigenvalue for this subblock given by
\begin{align}\label{lambda2proof}
	\Lambda_2 &\equiv {\rm Tr}\; \mathbf{v} \cdot\mathbf{v}^\dagger = \mathcal{P}_{A_1}\,.
\end{align}
In other words, the OSES has another entry corresponding to the purity of the subsystem $A_1$ that was not traced out.

\subsection{The particle's coherence extends across the bipartition}
\label{block3}
We are left with coherences between sites that are on opposite sides of the cut. This means that we deal here with matrix elements with indices $i_{1}\ldots i_m$ and $i_{m+1}\ldots i_L$ marking empty sites with a single position where $s=\pm$ can appear. Correspondingly, this block is of size $2m\times 2m$. 

we denote $G+_j$ as a string of $e$'s with a single $+$ at index $j$, and $G-_j$ as the same string but with a $-$ at index $j$. The most general form of such elements of $C$ is
\begin{align} \label{eq:coherenceblock}
			C_{Gs1_j ; Gs2_k} = \sum_{s3}\sum_{l=m+1}^{L} P_{j=s1 ; l=s3} P^*_{k=s2; l=s3}\,,
\end{align}
where the indices $s1,s2,s3=\pm$ and $1\leq j, k \leq m$. Taking into account that, in our single-particle case, $s3=-s1$ but also $s3=-s2$, we have nonvanishing elements solely when $s1=s2=s$. As a result, this block of $C$ splits into two independent subblocks, each of size $m\times m$.
		
Each of these $m\times m$ subblocks have two types of entries. On the diagonal of each subblock, $j=k$,
		\begin{align}
			C_{Gs_j, Gs_j} = \sum_{l=m+1}^{L} | P_{j=s ; l=-s} |^2\,.
		\end{align}
		The off-diagonal elements of each subblock, then, take the form
		\begin{align} \label{eq:coherenceblock2}
			C_{Gs_j, Gs_k} = \sum_{l=m+1}^{L} P_{j=s ; l=-s} P^*_{k=s; l=-s}\,,
		\end{align}
As in Sec.~\ref{block2}, we identify that each $s$-subblock is constructed as $\sum_{l=m+1}^{L}  \mathbf{v}_{s,l}\cdot \mathbf{v}_{s,l}^\dagger$ with $v_{s,l}$ being a column vector with entries $P_{j=s, l=-s}$, where $j=1\ldots m$. 
Because of the Hermiticity of the density matrix, these vectors are complex-conjugate to one another, i.e., $\mathbf{v}_{s,l} = \mathbf{v}_{-s,l}^{*}$.  Hence the two subblocks are identical and we expect to obtain $m$ two-fold degenerate eigenvalues contributing to the OSES from this case.

As in the previous two cases, we wish to relate these eigenvalues to a more physical intuition. To better understand the information encoded in the eigenvalues arising from the coherence block (Sec.~\ref{block3}), we can write each complex coefficient in Eq.~\eqref{eq:coherenceblock2} as $P_{j=s ; l=-s}=|P_{j=s ; l=-s}|e^{i\phi_{j,l,s}}$. Using Eq.~\eqref{coh_to_dens} we then write
\begin{align} \label{cond2}
		P_{j=s ; l=-s}=\left|\xi_{jl}\right|\sqrt{P_{j=f}P_{l=f}}e^{i\phi_{j,l}s}\,.
\end{align}
Plugging Eq.~\eqref{cond2} into Eq.~\eqref{eq:coherenceblock2}, we obtain
\begin{align*} 
   \resizebox{.9\hsize}{!}{ $C_{Gs_j, Gs_k} = \sum_{l=m+1}^{L} \left|\xi_{jl}\right|\left|\xi_{kl}\right|P_{l=f}\sqrt{P_{j=f}P_{k=f}}e^{i(\phi_{j,l}-\phi_{k,l})s}\,$}.
\end{align*}
Let us now consider the special case in which contributions to decoherence are such that $\xi_{jl}$ factorizes into $\xi_{j}\xi_{l}$. In this special case, $C_{Gs_j, Gs_k}$ takes the form
\begin{align*} 
\resizebox{.9\hsize}{!}{$ C_{Gs_j, Gs_k} = \left(\sum_{l=m+1}^{L} \left|\xi_{l}\right|^2 P_{l=f}\right)\xi_{j}\xi_{k}\sqrt{P_{j=f}P_{k=f}}e^{i(\tilde{\phi}_{j}-\tilde{\phi}_{k})s}\,$},
\end{align*}
where we have defined $\tilde{\phi}_{j}=\phi_{j,l}-\theta_j-\theta_l-\tilde{\phi}_{l}$ using the local decohorence contributions collected as $\xi_{j}=\left|\xi_{j}\right|e^{i\theta_{j}}$. Note, that $\tilde{\phi}_{j}$ is the phase associated with the amplitude of the pure state at position $j$.

The decoupling offered by the local decoherence assumption is quite illuminating: the sum over $l$ enters as a constant prefactor in all elements of the coherence block of $C$, cf.~Sec.~\ref{block3}. As a result, the subblock can be described, in similitude to (ii), as a product of a a single vector $\mathbf{u}$ with elements $\mathcal{A}\xi_{j}\sqrt{P_{j=f}}e^{i\tilde{\phi}_{j}}$ where we have defined $\mathcal{A}=\sqrt{\left(\sum_{l=m+1}^{L} \left|\xi_{l}\right|^2 P_{l=f}\right)}$. We, therefore, obtain in this special case that the coherence block contributes two degenerate OSES values solely. Their value is the squared norm of the vector, $\left|\mathbf{u}\right|^2$, and corresponds to a generalized cross-boundary concurrence, i.e., these values take the form
\begin{align} \label{eq:coherenceblock4vals}
	\Lambda_{3,4} = \left(\sum_{l=m+1}^{L} \left|\xi_{l}\right|^2 P_{l=f}\right)\left(\sum_{j=1}^{m} \left|\xi_{j}\right|^2 P_{j=f}\right)\,,
\end{align}
which is identically the squared norm of a vector made of concurrences between all pairs of two sites across the boundary, $\sqrt{P_{j=f}P_{l=f}}$, attenuated by local decoherence factors.
 
We conclude that under the local decoherence assumption the OSES of a single-particle will be composed of strictly four values regardless of the size of the system. 
It is important to note that the case of a pure state, $\xi_{ij}\equiv 1$ fits under this limiting assumption. 
Indeed, for a single-particle pure state case, the ES has only two values (see Appendix~\ref{app:singleparticleES}) and for pure states the OSES is just the tensor product of the ES with itself (c.f. Appendix~\ref{app:singleparticleES}), in conjunction with our result here of having four OSES values. Deviations away from this simple case will generate additional values arising from the interplay of dissipation and the state's coherence in the block discussed in Sec.~\ref{block3}. Nevertheless, these deviations are usually small and we commonly observe four dominant values in the single-particle case.
We have rigorously analyzed the single-particle case of the OSES. In the following two subsections we provide two intuitive examples to illustrate our result.

\section{Demonstrations}
\label{sec:demos}
\subsection{Example I: Charge-qubit Rabi oscillations}
As we have shown above, the OSES values contain important bipartite entanglement information. In order to demonstrate the physical relevance of these values, we will start with a simple case study of a charge qubit, i.e., a single-excitation hopping between two sites. The Hamiltonian describing such a setup simply is
\begin{align}
H = -t c^\dagger_1 c^{\phantom{\dagger}}_2 + \rm{h.c.}	\,,
\end{align}
where $c_1$ and $c_2$ are the single-particle annihilation operators on sites 1 and 2, respectively. We have assumed a vanishing on-site potential and use $t$ as the hopping amplitude. 

The corresponding $4\times 4$ density matrix $\rho(t)$ describing state at any time $t$ only has a single $2\times 2$ non-vanishing subblock corresponding to the single particle sector:
\begin{equation}
\rho^(t) = \left( \begin{array}{cc}
	\rho_{11}(t) & \rho_{12}(t)\\
	\rho_{12}^*(t) & \rho_{22}(t) 
	\end{array} \right)\,.
\end{equation}
Coupling the system to local baths that interact capacitively with the sites, we describe the system dynamics using a standard master equation in Lindblad form
\begin{equation} \label{eq:masterEquation}
    \partial_t \rho = -i \left[ H, \rho \right] + \gamma \sum_i \left( 2 \hat{n}^{\phantom{\dagger}}_i \rho \, \hat{n}_i^\dagger - \{ \hat{n}_i^\dagger  \hat{n}^{\phantom{\dagger}}_i, \rho \} \right)\,,
\end{equation}
where $\hat{n}_i$ is a local density operator at site $i$, and $\gamma$ is the coupling strength to the (infinite temperature) bath.

For this simple case, one can easily identify the nonvanishing coefficients $P_{1=f}=\rho_{11}(t)$, $P_{2=f}=\rho_{22}(t)$, and $P_{1=+,2=-}=\rho_{12}(t)$. There is only one possible spatial bipartition and the resulting $C$-matrix is already diagonal with the following OSES:
\begin{align}
\label{lamd1}
\Lambda_1=&\rho_{11}^2(t)\equiv n_1(t)^2\,,\\
\label{lamd2}
\Lambda_2=&\rho_{22}^2(t)\equiv n_2(t)^2\,,\\
\label{lamd3}
\Lambda_{3,4}=&\left|\rho_{12}(t)\right|^2 = \left|\xi_{12}(t)\right|^2 n_1(t)n_2(t)\,,
\end{align}
where $n_i$ is the density at site $i$. 
Above, we have used Eq.~\eqref{coh_to_dens} to relate the off-diagonal density matrix elements to the diagonal ones using the decoherence prefactor $\xi_{12}(t)$. 
Consequently, while $\Lambda_1$ and $\Lambda_2$ follow trivially the square of the densities at the two sites (which are indeed the purity of the subsytems), the values $\Lambda_{3,4}$ are degenerate and correspond to the square-root of the quantum state's concurrence.

\begin{figure}[t!]
\begin{center}
\includegraphics[width=\columnwidth]{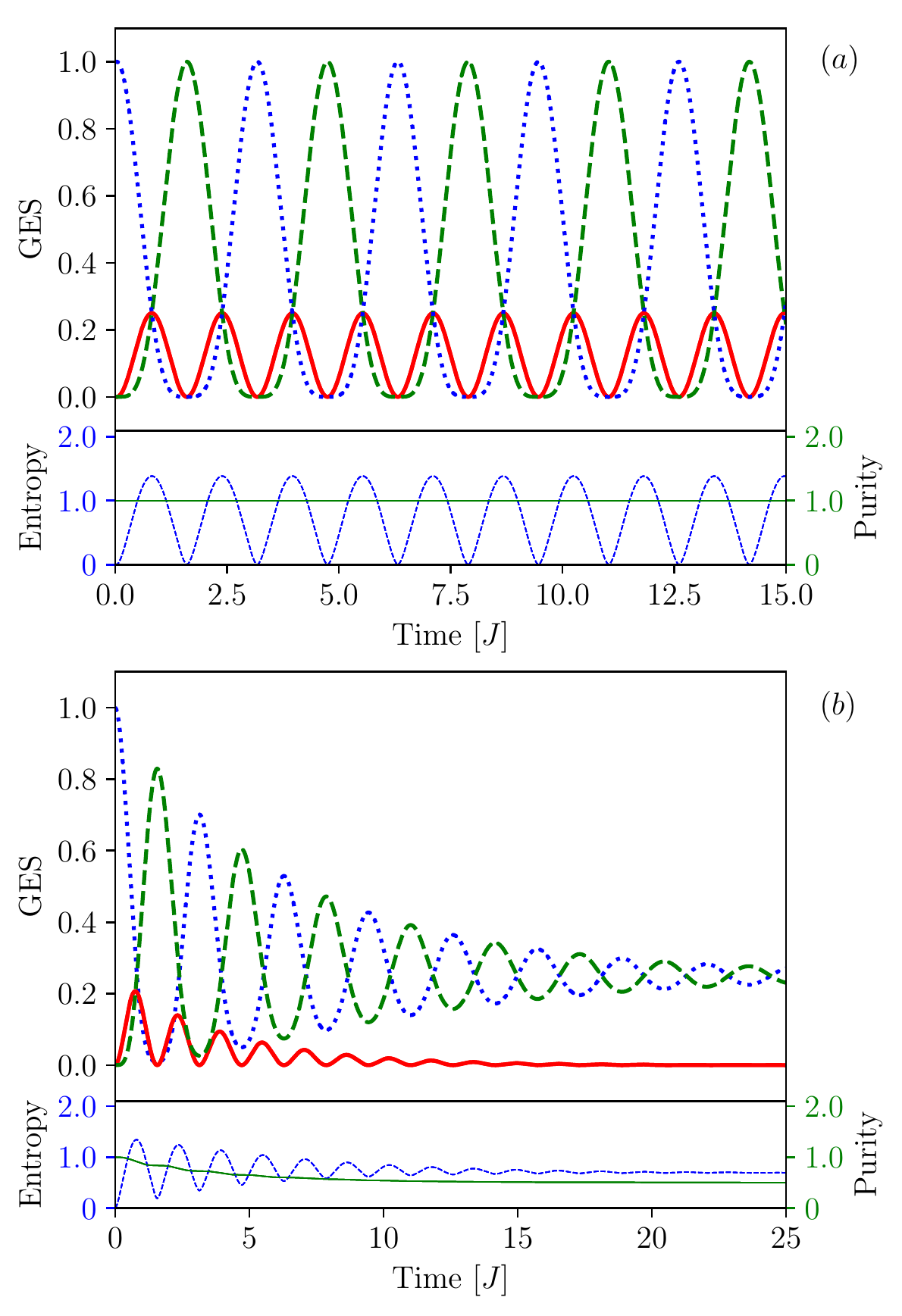}
\end{center}
\caption{(Color online) Plots of the OSES values (top) and respective measures (bottom) for a two-site charge-qubit case as a function of time, obtained through MPS simulations (see Appendix~\ref{app:MPS}). The particle is initialized on site 1. The OSES values $\Lambda_1$, $\Lambda_2$ are depicted as dotted (blue) and dashed (green) lines [cf.~Eqs.~\eqref{lamd1} and \eqref{lamd2}]. The values $\Lambda_{3,4}$ are depicted as solid (red) lines [cf.~Eq.~\eqref{lamd3}]. The resulting operator space entanglement entropy and purity measures are depicted as dashed (blue) and solid (green) lines.  (a) The $\gamma=0$ case. (b) The $\gamma=0.25$ case.}
\label{fig1}
\end{figure}

In Fig.~\ref{fig1}(a), we plot the $\Lambda_i$ for a pure state evolution ($\gamma=0$). Whenever the particle is equally spread over the sites, the OSES values are degenerate, whereas mostly they have a ``1,1,2''-degeneracy structure corresponding to squared-densities and a doubly-degenerate concurrence value. The lower panel shows the respective operator space entanglement entropy and total purity computed from the OSES.
In Fig.~\ref{fig1}(b), we present the obtained OSES and respective measures for non-Hamiltonian evolution (c.f. \eqref{eq:masterEquation}, with $\gamma=0.25$). Over time, the Rabi oscillations reduce alongside a decaying concurrence towards an equal weight statistical mixture. This can also be seen in a reduction with time $t\rightarrow \infty$ of the entanglement entropy saturates at $\log 2$ and purity at $1/2$, as expected.

It is clear that for such a simple case the four OSES values contain all the information on the single-particle density matrix on two-sites, since there are merely four elements in the density matrix in total. This simple example, however, points to a natural understanding of the bipartite information that the OSES contains, namely, we have two values that contain weights on how much of the particle is to either side of the bond and two degenerate values that correspond to the amount of cross-bond coherence.

\subsection{Example II: Four-site case}
We now extend our demonstration to the case of a single-particle hopping across four sites. The Hamiltonian keeps its simple structure, i.e.
\begin{align}
H_{\rm cq} = -t \sum_{i=1}^3 c^\dagger_i c^{\phantom{\dagger}}_{i+1} + \rm{h.c.}	\,,
\end{align}
with a corresponding density matrix $\rho(t)$ of size $2^4 \times 2^4$ having  an effective $4 \times 4$ non-vanishing single-particle subblock. 
Also in this case we couple a dephasing bath to the sites of the system, c.f. Eq.~\ref{eq:masterEquation}.
 
\begin{figure}[t!]
\begin{center}
\includegraphics[width=\columnwidth]{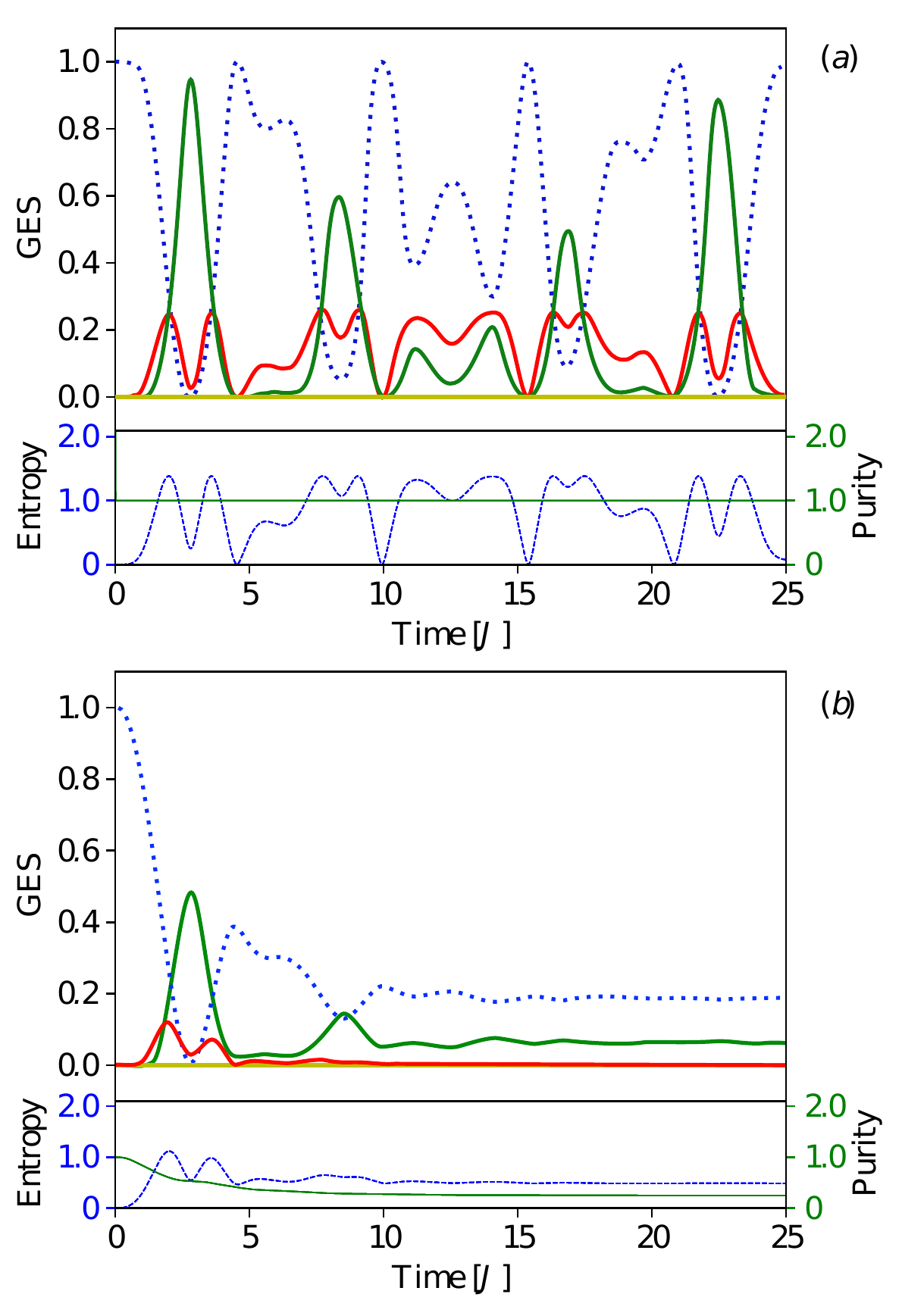}
\end{center}
\caption{(Color online) Plots of the OSES values (top) and respective measures (bottom) for a single-particle on four-site case as a function of time. The particle is initialized on site 1, and we display the central bond OSES. The values $\Lambda_1$, $\Lambda_2$ are depicted as dotted (blue) and dashed (green) lines [cf.~Eq.~\eqref{lamd1foursite} and \eqref{lamd2foursite}]. The OSES values $\Lambda_{3,4}$ and $\Lambda_{5,6}$ are depicted as solid (red and yellow, respectively) lines. The resulting generalized entanglement entropy and parity measures are depicted as dashed (blue) and solid (green) lines.  (a) The $\gamma=0$ case. (b) The $\gamma=0.3$ case.}
\label{fig2}
\end{figure}

In this case, the excitation has a chance to develop a full coherent spread to either side of the central bond as well as cross-bond coherence. 
In other words, there are two sites to each side of the central bond, each allowing for a charge qubit. 
We focus here on the OSES of the central bond, and by the construction of section~\ref{sec:singleexcitationOSES} we have:
\begin{align}
\label{lamd1foursite}
\Lambda_1=&n_1^2 + 2|\xi_{12}|^2 n_1 n_2 + n_2^2\,\\
\label{lamd2foursite}
\Lambda_2=&n_3^2 + 2|\xi_{34}|^2 n_3 n_4 + n_4^2 \,
\end{align}
which correspond to the purities of both subsystems. In this case, there are two sets of doubly-degenerate pairs of OSES entries, $\Lambda_{3,4}$ and $\Lambda_{5,6}$, corresponding to cross-bond coherence. If the coherence factors $\xi_{ij}$ were to factorize to $\xi_{ij} = \xi_i \xi_j$ as in section~\ref{block3}, only two degenerate cross-boundary generalized concurrence values would remain:
\begin{align}
\label{lamd3foursite}
\Lambda_{3,4}= \xi_1\xi_3 n_1 n_3 + \xi_1\xi_4 n_1 n_4 + \xi_2\xi_3 n_2 n_3 + \xi_2\xi_4 n_2 n_4\,.
\end{align}
Due to our local coupling to the baths, we indeed find (c.f. Fig.~\ref{fig2}) that $\Lambda_{5,6}$ are generically much smaller than the other entries and appear as a perturbative deviation.

In Fig.~\ref{fig2}(a), we plot $\Lambda_i$ for a pure state evolution ($\gamma=0$). We see the coherent quantum walk of the particle between the four sites reflected in the OSES. The bipartition yields an effective qubit decomposition with weights to each side of the bond: (i) $\Lambda_1$ and $\Lambda_2$ correspond in the pure case to the square of the total density to the left and right of the bond, (ii) $\Lambda_{3,4}$ are a generalized concurrence across the bond, and (iii) $\Lambda_{5,6}\equiv 0$. In the lower part, the respective generalized entanglement entropy and total purity are computed from the OSES.
In Fig.~\ref{fig2}(b), the $\gamma=0.3$ case is depicted. From the OSES, we see that the local purities to each side of the bond decay to
$\sqrt{1/8}$, symbolizing a homogeneous delocalization of the particle across the system, i.e., $n_i=1/4$. The cross-bond coherence is dominated by two main values that decay over time. In the limit $t\rightarrow \infty$ the operator space entanglement entropy goes to $(3/4)\log 2$ and purity saturates at $1/4$, as expected.

\section{Conclusion and outlook}
\label{sec:outlook}
In this work we study the bipartitioning of a mixed state density matrix, and obtain the eigenspectrum of the resulting reduced operator. Our procedure is analogous to that used for pure states, where the obtained eigenspectrum is referred to as the entanglement spectrum~\cite{Li2008}. We obtain an operator space entanglement spectrum (OSES) that can be related to physical properties of the system. In particular, the OSES contains information on (i) the total purity of the system, (ii) the purity of the two subsystems, as well as (iii) the cross-boundary coherence. 

For the case of a mixed state density matrix with a single excitation, the bipartitioning can be performed analytically. We consider an important limiting case of \textit{local dissipation}. This local dissipation ansatz directly matches both the pure state case as well as the fully decohered density matrix scenario, where only four values are sufficient to describe the state. 
The obtained OSES in intermediate cases of partially decohered states can be thought of perturbative corrections away from this point.

Whereas in the single-particle case the $C$-matrix \eqref{eq:C} could be decomposed into three blocks, in the many-body case there will be many more blocks to be considered. Nonetheless, we expect similarities to persist. 
For example, the block corresponding to all particles on the left of the bipartitioning will always correspond to the purity of that subsystem. 
Hence, as matrix product decomposition allows for a straightforward numerical way of obtaining the OSES also in many-body cases, the knowledge of what the values encode (even without their exact analytical derivation), is then useful for simulating many-body mixed state dynamics. 
Unitary transformations of the basis before truncation can then help preserve these properties~\cite{White2017}.

Last, in our analysis, we rely on the fact that the particle-number of the system is conserved in order to find a meaningful block decomposition of the $C$-matrix in Eq.~\eqref{eq:C}. Hence, the OSES of systems where couplings between particle sectors occur are not covered in this work. 

\acknowledgments 
 We would like to thank C.~Carish, R.~Chitra, M.~Ferguson, M.~Fischer, S.~Huber and C.D.~White for useful discussions on this work. We acknowledge financial support from the Swiss National Science Foundation (SNSF).

%

\appendix
\section{MPS implementation}
\label{app:MPS}
A practical way of obtaining the entanglement spectrum is through the use of a matrix product state (MPS) decomposition. In MPS simulations, the coefficients of the wavefunction in a given basis $|i_1 \ldots i_L\rangle$ are replaced by a product over local matrices (for extensive reviews on the topic of MPS, we point the reader to~\cite{Sch11a, Oru14a}):
\begin{align}
	|\psi\rangle_{i_1\ldots} = \text{Tr}\left[ \Gamma^{[1]}_{i_{1}}\Lambda^{[1]} \dots \Gamma^{[L]}_{i_{L}}\Lambda^{[1]} \right] |i_1\ldots i_L\rangle \text{.}
\label{eq:mps}
\end{align}
In this representation, the diagonal matrix $\Lambda^{[m]}$ contains the entanglement spectrum values for a bipartitioning between sites $m$ and $m+1$. The dimension of the matrices (called the bond dimension) is what limits the number of \emph{independent} coefficients that can be constructed, and hence controls the level of the approximation of the MPS~\cite{Hastings07a, Hastings10a, Schuch08a}. 
We remark that such a representation of an arbitrary wavefunction can be exact as long as the size of the local matrices is chosen large enough~\cite{Vid03a}. 
In the current case of our single-particle analysis, the use of MPS methods is uncalled for. 
They provide, however, a straightforward extension to the multi-particle case.

Algorithms such as DMRG or TEBD are designed to find the set of $\Gamma_{i_n}$ and $\Lambda^{[n]}$ matrices for which the MPS best approximates the actual wavefunction of the system. 
Instead of considering pure states, we may construct an MPS of the vectorized density matrix $\ket{\rho}$ of Eq.~\eqref{eq:fanoForm} as in Ref.~\cite{Zwolak2004}. 
The corresponding $\Lambda^{[n]}$ matrices then correspond to the Schmidt values of $Q$ when bipartitioned at $n$, which themselves are equal to the square roots of the eigenvalues $\Lambda_i$ of $Q_{A_1}$. 
These values can be directly interpreted as the OSES, with the additional remark that in standard TEBD algorithms they are typically normalized such that $\sum_i \Lambda_i^2 = 1$. 
In the above, we have taken out this normalization factor so that instead this sum equals the purity of the state. For the computation of any observables this normalization factor should of course be included.

\section{Single-particle entanglement spectrum}
\label{app:singleparticleES}
In this Appendix we derive the expressions for the entanglement spectrum in 1D systems containing a single-particle. It serves as a step to the full proof in the case of the mixed state.

Consider a system of $L$ sites, where the local basis states are $|0\rangle$ and $|1\rangle$. Let us denote the first $L/2$ sites as subsystem $A$ and the last $L/2$ sites as subsystem $B$. We introduce the single particle basis states for each of the halves as
\begin{align}
	|n\rangle_A = \bigotimes_{i=0}^{L/2-1} |\delta_{in}\rangle, \quad \text{and} \quad |n\rangle_B = \bigotimes_{i=L/2}^{L} |\delta_{in}\rangle.
\end{align}
The full single particle basis states can then be spanned by the states 
\begin{align}
\label{eq:fullspstate}
	|n \rangle = \theta \left(n < \frac{L}{2}\right) |n\rangle_A \otimes |\mathbf{0}\rangle_B + \theta \left(n \geq \frac{L}{2}\right) |\mathbf{0}\rangle_A \otimes |n\rangle_B,
\end{align}
where $|\mathbf{0}\rangle$ denotes a tensorproduct of only the $|0\rangle$ basis states on the respective subsystems. Even though the expression in Eq.~\eqref{eq:fullspstate} looks like a superposition, these basis states are all pure product states due to the Heaviside $\theta$-function. 

The partial trace over subsystem $B$ of the density matrix $\rho = \sum_{n,m} |n\rangle\langle m|$ can now be performed as
\begin{widetext}
\begin{align}
	\text{Tr}_B \rho &= \sum_{\alpha\, n\, m} {}_B\langle \alpha | \rho | \alpha\rangle_B + {}_B\langle \mathbf{0} | \rho | \mathbf{0} \rangle_B \\
				&= \sum_{\alpha\, n\, m} c_n c_m^*  \Big[ \theta \left(n,m \geq \frac{L}{2}\right)   |\mathbf{0}\rangle_A \langle \mathbf{0} |  \langle \alpha | n \rangle_B  \langle m | \alpha \rangle_B +  \theta \left(n,m < \frac{L}{2}\right) |n\rangle_A \langle m| \Big] \\
				&= \sum_{\alpha=L/2}^{L} |c_\alpha|^2 |\mathbf{0}\rangle_A \langle \mathbf{0} | + \sum_{i=0}^{L/2-1} |c_i|^2 |\psi\rangle_A \langle \psi |,
\end{align}
\end{widetext}
where in the last step, $| \psi \rangle $ is a normalized wavefunction corresponding to $\sum_{i\in A} c_i |i\rangle_A$.
Hence the partial traced density matrix consists of two blocks, each with a single eigenvalue. These eigenvalues, $\lambda_1 = \sum_{\alpha=L/2}^L |c_\alpha|^2$ and $\lambda_2 = \sum_{i=0}^{L/2-1} |c_i|^2$, correspond to the total density of subsystems $B$ and $A$ respectively. Notably, for the pure state case the partial-trace procedure generates two independent subblocks. For the mixed case, an extra subblock will appear that corresponds to possible coherences of the particle across the bipartioning.

\subsection{Relation to OSES}
\label{app:ESvsGES}
For a pure state, we compare the MPS coefficients $\text{Tr}\left[ \Gamma^\sharp_{i_{1}} \lambda^{[1]\sharp} \dots \Gamma^\sharp_{i_{L}} \lambda^{[L]\sharp} \right]$ of a density matrix $\ket{\rho}$ to those obtained from $\ket{\Psi}\bra{\Psi}$:

\begin{widetext}
\begin{align}
 \ket{\Psi}\bra{\Psi} &= \sum_{\{i_{1}\ldots i_{L}\}} \sum_{\{j_{1}\ldots j_{L}\}}
	\text{Tr}\left[ \Gamma_{i_{1}} \lambda^{[1]} \dots \Gamma_{i_{L}} \lambda^{[L]} \right]\, \text{Tr}\left[ \Gamma^*_{j_{1}} \lambda^{[1]} \dots \Gamma^*_{j_{L}} \lambda^{[L]} \right]| i_{1}\dots i_{L} \rangle \langle j_{1}\dots j_{L}| \nonumber \\ 
	&= \sum_{\{i_{1}\ldots i_{L}\}} \text{Tr}\left[ \left(\sum_{j_1}\Gamma_{i_{1}}\otimes \Gamma^*_{j_{1}}\right) \lambda^{[1]}\otimes\lambda^{[1]} \dots  \Bigg(\sum_{j_L}\Gamma_{i_{L}}\otimes \Gamma^{*}_{j_{L}}\Bigg) \lambda^{[L]}\otimes\lambda^{[L]} \right] | i_{1}\dots i_{L} \rangle \langle j_{1}\dots j_{L}|
\end{align}
\end{widetext}
Via a basis transformation, we may relate $|i_1 \rangle\langle j_1 |$ to the $\sigma$ matrices. This transformation will only redefine the $\Gamma$ matrices however, whereas for the $\lambda$ matrices we immediately find the relation $\lambda^{[n]\sharp} = \lambda^{[n]}\otimes \lambda^{[n]}$. This relationship also directly leads to the generalized entanglement entropy for a pure state being twice the entanglement entropy, since $\ln \lambda \otimes \lambda = \ln \lambda \otimes 1 + 1 \otimes \ln \lambda$, and hence
\begin{align}
	\text{Tr} \lambda \otimes \lambda \ln \lambda \otimes \lambda &= \text{Tr} (\lambda \ln \lambda) \otimes \lambda + \lambda \otimes (\lambda \ln \lambda) \\
	&= 2\; \text{Tr} (\lambda \ln \lambda ) \text{Tr} \lambda \\
	&= 2\; \text{Tr} \lambda \ln \lambda .
\end{align}

\end{document}